# Effect of Indium doping on structural and thermoelectric properties of SnTe

Diptasikha Das[1][0009-0000-8144-3441], A. Jana[2], S. Mahakal[2], Pallabi Sardar[1], J. Seal[1], Shamima Hussain[3] and Kartick Malik*[2][0000-0002-9175-3467]

[1]Department of Physics, Adamas University, Kolkata, West Bengal, India
[2]Department of Physics, Vidyasagar Metropolitan College, Kolkata, West Bengal
[3]UGC-DAE Consortium for Scientific Research, Kalpakkam Node, Tamil Nadu, India.
*Corresponding Author Email: kartick.phy09@gmail.com

**Abstract.** The solid state reaction method is employed to synthesize $Sn_{1-x}In_xTe$ (x=0.00, 0.02, 0.04, and 0.05) samples. Power Factors of synthesized samples are estimated from resistivity ($\rho$) and thermopower (S) data. Modifications in structural parameters, $\rho$, and S owing to In doping in SnTe thermoelectric (TE) material are reported. In-depth structural analysis, employing Rietveld refinement of X-ray diffraction (XRD) data, confirms the substitution of Sn by In. A minute amount of embedded phases in synthesized samples is revealed from the refinement of XRD data. Williamson-Hall and modified Williamson-Hall methods are employed to estimate dislocation density and strain. The highest power factor and maximum host phases are simultaneously achieved for the $Sn_{0.96}In_{0.04}Te$ sample amid the synthesized $Sn_{1-x}In_xTe$ samples.

## 1 Introduction

A worldwide resurgence is going on to get rid of the trap of fossil fuel. Fossil fuel is one of the primary sources employed for the conversion of heat to electricity using available compatible technologies. However, uses of natural gas, fuel, and coal to generate electricity have many disadvantages and adverse effects on atmospheric pollution and global warming etc. Nowadays, the development of renewable energy is one of the important and pertinent topics in research for sustainable development. Almost 60% of the heat energy is wasted during the conversion of heat to electricity using fossil fuel and conventional technologies. The development of sustainable technologies for effective use, conversion, and recycling the waste energy is another way to solve the problem. Affirmative action may be taken for the recovery of waste heat by employing Thermoelectric (TE) technology [1, 2]. TE technology is a potential alternative to recover a huge amount of waste heat, without adverse impact on the environment. TE device is a solid-state energy converter for the conversion of heat to electrical energy and vice versa without any mechanical moving parts [1-3]. TE materials are a special type of narrow band gap semiconductor. The efficiency of a TE material is characterized by a term, Figure of Merit, $ZT = \left(\dfrac{S^2}{\rho\kappa}\right)T$ where S, ρ, and κ are the Seebeck coefficient, electrical resistivity, and thermal conductivity of the material, respectively [1-3]. $S^2/\rho$ is known as Power Factor (PF) and also indicates the maximum derivable output power from the TE device.

TE materials are categorized as low-temperature (temperature < 300 K), mid-temperature (temperature 500 K to 800 K), and high-temperature (temperature > 900 K) according to the performance or highest ZT obtained in the temperature range. Mid-temperature TE materials have the potential to recover waste heat in industry and car exhaust. Some well-known mid-temperature TE materials are Lead chalcogenide, Tin chalcogenide, filled Skutterudite, and Half-Heusler based TE materials, etc [4-7]. Lead telluride (PbTe) and its alloys have been considered as a primary mid-temperature TE material for industry-based systems [5]. However, PbTe has limited application in daily life due to the toxicity of Pb. An alternative lead-free compound, SnTe, has been introduced as a potential TE material in the mid-temperature range. Environment-friendly, SnTe is also crystallized as a rock salt structure, and the electronic structure of SnTe consists of multiple valence bands (light-hole, L and heavy-hole, Σ bands) alike PbTe [8]. Electronic band gap ($E_g$), i.e., gap between conduction band and L band of pristine SnTe ($E_g \sim 0.18$ eV), is very small compared to PbTe ($E_g \sim 0.30$ eV) [8]. However, the gap between the two valence bands, i.e., L and Σ of SnTe ($\Delta E_{L-\Sigma} \sim 0.35\ eV$), is large concerning PbTe ($\Delta E_{L-\Sigma} \sim 0.17\ eV$) [8]. Intrinsic SnTe possesses a high hole concentration (n) ($10^{20}$ - $10^{21}$ cm$^{-3}$) relative to PbTe due to inherent Sn vacancies [8]. High hole concentration and dissimilar electronic band gap are the inherent reasons for very low S and high $\kappa_e$ in SnTe, compared to PbTe [9]. Hence, pristine SnTe exhibits very poor ZT. However, the PF of SnTe may be enhanced by optimization of *n* and band engineering method via valence band convergence, resonance level engineering, synergistic effect, and increasing carrier effective mass [10,11]. Furthermore, the $\kappa_L$ may be reduced to optimize ZT by introducing various types of crystal defects, viz., point defects, dislocations, interfaces, and precipitates in the bulk matrix. Crystal defects owing to nanostructure are a very useful method to reduce $\kappa_L$ [10,11]. Nano-sized SnTe, synthesized by the hydrothermal method, shows low $\kappa_L \sim 0.6\ wm^{-1}K^{-1}$ and high ZT~0.49 at 803 K [12]. ZT~1.1 at 873 K has been obtained for nano-structured In$_{0.0025}$Sn$_{0.9975}$Te [13]. ZT ~ 1.3 at 873 K has been obtained for nano-composite SnCd(0.03)Te + 2% CdS/ZnS, synthesized by the melting process [9].

The effect of embedded phases and In doping on structural and transport properties of the synthesized samples has been investigated in this article. Rietveld refinement is employed to carry out in-depth structural analysis of Sn$_{1-x}$In$_x$Te (x=0.00, 0.02, 0.04, and 0.05) samples. X-ray diffraction and Rietveld refinement confirm a decrease in lattice parameter with increasing In concentration and concomitant replacement of Sn by In in the SnTe matrix. Mix-phase Rietveld refinement supports the presence of embedded phases in the host matrix. Variation in room temperature resistivity (ρ) and thermopower (S) is corroborated with structural data. The highest power factor is obtained for the Sn$_{0.96}$In$_{0.04}$Te synthesized sample. It is crucial to note that the maxi-mum host phase is obtained for the $Sn_{0.96}In_{0.04}Te$ synthesized sample, as revealed from the structural analysis.

## 2      Experimental Details

$Sn_{1-x}In_xTe$ (x=0.00, 0.02, 0.04, and 0.05) samples were synthesized by the solid state reaction method. Stoichiometric amounts of elements (Sn, In, and Te) were loaded in a quartz tube and vacuum sealed at $10^{-3}$ Pa to avoid oxidation. Vacuum-sealed quartz ampoules were kept in the furnace, and the temperature was raised to 973 °C within 5 h. It was then sintered at 973 °C for 15 h to homogenize the alloys, followed by ice quenching to avoid the intermediate phase segregation. In-depth structural characterizations were performed by the Rietveld refinement of X-ray diffraction (XRD) data employing FullProf software.

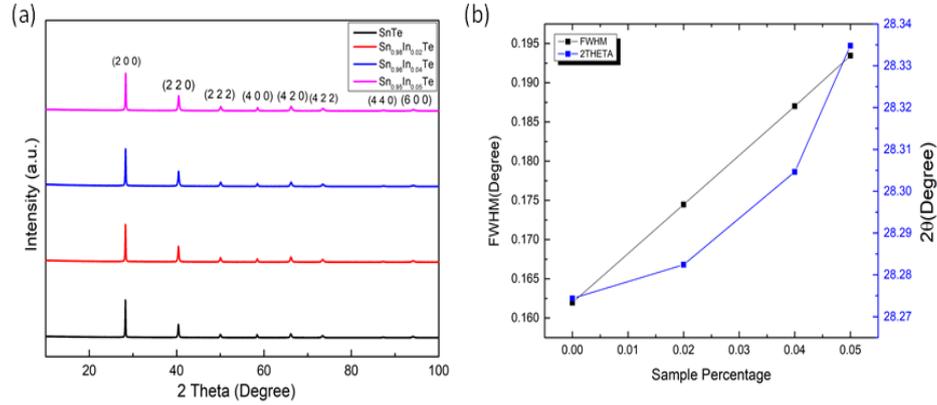

**Fig.1.** (a) X-ray diffraction pattern of $Sn_{1-x}In_xTe$ (x=0.00, 0.02, 0.04, and 0.05) and (b) Variation of Full width at half maxima (FWHM) and position (2θ) of the most intense peak (200) for different In concentration (x=0.00, 0.02, 0.04, and 0.05).

## 3      RESULTS AND DISCUSSION

XRD patterns of the synthesized samples are presented in Fig. 1(a). XRD data confirms that the synthesized samples, $Sn_{1-x}In_xTe$, are crystallized as a single phase and diffraction peaks are indexed with space group $Fm\bar{3}m$. FWHM of the most intense peak (200) gradually increases with In concentration, indicating degradation of crystal quality and distortion in the structure (Fig. 1(b)). An increase in peak position for (200) peaks for the synthesized samples signifies a decrease in lattice parameter (Fig. 1(b)). The atomic radius of Sn is greater than In; replacement of Sn by In causes a decrease in lattice parameter, and reflected in the XRD pattern. However, in-depth structural analysis by Rietveld refinement also confirms the decrease in lattice parameters (Fig. 2(a)). Mix-phase Rietveld refinement indicates the presence of minute amounts of embedded phases in the matrix of the host phases. Williamson-Hall and modified Williamson-Hall plots are performed to estimate strain and dislocation density, respectively. The variation in dislocation density and strain with In concentration is represented in Fig. 2(b). Dislocation density of the synthesized samples is estimated in comparison to SnTe (x=0.0). Embedded phases, foreign elements, and In incorporation in the SnTe matrix cause an increase in strain and dislocation density (Fig. 2(b)). In order to get an estimate of PF, room temperature $\rho$ and S measurements of $Sn_{1-x}In_xTe$ synthesized samples are carried out. The variation of room temperature $\rho$ and S as a function of In concentrations is plotted in Fig. 2(c). It is crucial to note that the variation of $\rho$ and S is contradictory; $\rho$ decreases but S increases. It is purely related to the variation of the double valence bands and the position of Fermi surface. PF is estimated from the room temperature $\rho$ and S, and presented in Fig. 2(d). PF of the 4% In doped sample is maximum among the synthesized $Sn_{1-x}In_xTe$ (x=0.00, 0.02, 0.04, and 0.05) samples.

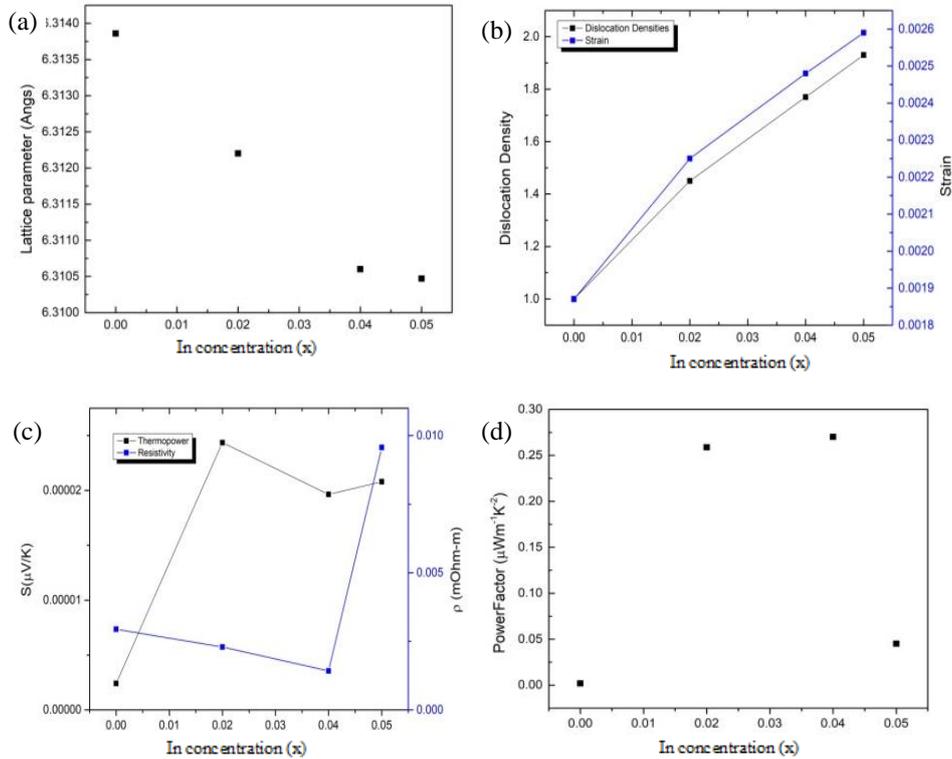

**Fig. 2.** (a) Lattice parameters of synthesized samples, $Sn_{1-x}In_xTe$ (x=0.00, 0.02, 0.04, and 0.05) as obtained from Rietveld refinement. (b) Dislocation density of $Sn_{1-x}In_xTe$ (x=0.00, 0.02, 0.04, and 0.05) samples with respect to SnTe. (c) Room temperature S and $\rho$ for different In concentration (x=0.00, 0.02, 0.04, and 0.05) (d) Indium concentration dependent PF of the synthesized $Sn_{1-x}In_xTe$ (x=0.00, 0.02, 0.04, and 0.05) samples.

## 4 SUMMARY

The solid state reaction method is employed to synthesize $Sn_{1-x}In_xTe$ samples. In-depth structural analysis is performed using XRD data employing Rietveld refinement, confirming the incorporation of In at the position of Sn. Embedded phases in the host matrix, i.e., SnTe, are revealed from the refinement. An increase in dislocation density and strain is related to the embedded phases. In doping and embedded phases cause a resonance level and shift in the Fermi surface in the SnTe host matrix. It is noteworthy to mention that 4% In doped sample shows maximum S and maximum host phases among the synthesized In doped SnTe samples. Room temperature structural and transport properties are corroborated in this endeavor.

**Acknowledgments**

This work is supported by the UGC-DAE-CSR Kalpakkam, India (CRS/2022-23/04/893) in the form of a sanctioned research project.